
Figures not included. Contact the authors.

\input jnl.tex
\input reforder.tex
\input eqnorder
\tolerance 1500
\baselineskip=0.8truecm
\def\line{\hbox to \hsize} \line{\hfil November 5, 1992}
\vglue 0.2truecm

\centerline{\bf DYNAMICAL PROPERTIES}
\bigskip
\centerline{\bf OF A SINGLE HOLE IN AN ANTIFERROMAGNET}

\vskip 1.9truecm

\centerline {Didier POILBLANC,$^{1,2}$}
\smallskip
\centerline{Timothy ZIMAN$^{1,}$\plainfootnote{$^\dagger$}{On leave from:
Department of Physics and Astronomy,
University of Delaware, Newark, DE 19716}, H. J. SCHULZ,$^2$}
\smallskip
\centerline{Elbio DAGOTTO$^{1,}$\plainfootnote{$^\ddag$}{Permanent address:
Department of Physics,
Florida State University, Tallahassee, FL 32306}}
\vskip 1.7truecm
\centerline{$^1$Laboratoire de Physique Quantique\plainfootnote{$^*$}{Unit\'e
Associ\'ee au Centre National de
    la Recherche Scientifique}}
\smallskip
\centerline{Universit\'e Paul Sabatier}
\smallskip
\centerline{31062 Toulouse, France}
\vskip 0.45truecm
\centerline{and}
\vskip 0.45truecm
 \centerline{$^2$Laboratoire de Physique des Solides{$^*$}}
\smallskip
\centerline{Universit\'e Paris-Sud}
\smallskip
\centerline{91405 Orsay, France}

\newpage

{\bf Abstract:}
A finite size scaling analysis of the spectral function and
of the optical conductivity
of a single hole moving in an antiferromagnetic background
is performed. It is shown that both the low energy quasiparticle peak
and the broad higher energy structure are robust with
increasing cluster size from $4\times 4$ to $\sqrt{26}\times\sqrt{26}$
sites. In the abscence of spin fluctuations, for most static or dynamical
quantities saturation occurs when the size exceeds a characteristic
size $N_c(J_z)$. Typically, 16 and 26 site clusters give reliable
results for $J_z>0.75$ and $J_z>0.3$
respectively. The hole
optical mass is shown to be very large ($>20$)
in agreement with the small bandwidth. Due to the energy gap to flip a spin in
the vicinity of a hole, a small gap $\propto J_z$ separates the low energy
delta-function from the rest of the spectrum
in the dynamical correlation functions.
On the other hand, with $J_\perp$
this gap seems to disappear with increasing system size as one would expect
since the spin waves are gapless in the thermodynamic limit.
The large momentum dependence of the quasiparticle
weight in the isotropic case is inconsistent with a string picture
but agrees well with the self-consistent Born approximation.
An accurate estimation of the higher energy part of the spectral
functions of the t--J model can be made for
momenta close to $(0,0)$ or $(\pi,\pi)$;
it consists of a broad featureless band.
For momenta close to the FS, no evidence for excited
string states is found. This may be either because the string scenario
breaks down or because exact
calculations at momenta near $(\pi/2,\pi/2)$ are limited by the differing
geometries of the finite clusters.
In the Ising limit, the spectral function depends also crucially on the
momentum but a feature resembling a string excitation is seen for momenta
close to $(0,0)$.
Our data on the finite frequency optical conductivity of
a single hole in the t--J model
reveal that the total spectral weight at finite frequency has a strong
size dependence and {\it decreases} with increasing size (for J not too small).
This suggests that the optical mass and the optical
absorption of the t--J model may have been overestimated in the past.

\vskip 0.5truecm

\noindent PACS Indices: 75.10.Jm, 75.40.Mg, 74.20.-z

\endtitlepage

\vfill\eject       

\centerline{\bf I. INTRODUCTION}

The recent discovery of high-$T_c$ superconducting copper oxides has
motivated a considerable
theoretical effort in the field of strongly correlated
fermions on two-dimensional lattices.
One of the central issues connected to the
problem of superconductivity is whether these systems behave as ordinary
Landau-Fermi liquids or whether they provide the first experimental
realization of a two-dimensional (2D) marginal fermi liquid.\refto{varma}
Besides the well-known Nagaoka
theorem,\refto{nagaoka} almost no other
exact results exist for this problem
in 2D. The validity of perturbative expansions,
as well as other analytic approximations,
remains controversial. Therefore in the last five years
a large effort in developing numerical methods
has been made.\refto{review}
Based on the Lanczos algorithm, exact diagonalization (ED)
techniques have proven to be very successful in
investigating both static and dynamical properties of strongly
correlated Hamiltonians specially in one and
two dimensions. Contrary to quantum
Monte Carlo (QMC) methods, ED does not present ``sign-problems'' or
other instabilities. Dynamical correlations are easily obtained
using this technique.
However, ED has been restricted so far to small systems,
typically $4\times 4$ clusters, due to the exponentially fast growth
of the Hilbert space with the system size.
Furthermore, finite size scaling analysis have not been
carried out in a systematic way in doped systems.
The goal of this paper is
to present such a finite size scaling analysis
in the case of a single hole moving in an
antiferromagnetic background. This paper is a development
of previous work.\refto{scaling1,scaling2}

In spite of its apparent simplicity,
the problem of a single hole in an antiferromagnetic background
is rather complicated.
It addresses many fundamental issues in the physics of
strongly correlated fermions. In particular, the interplay between
antiferromagnetic long range order (LRO) and charge degrees of freedom
is of particular interest. Previous numerical\refto{dyna,spectralnum,scaling2}
and analytical work\refto{spectralana}
suggested that the hole, dressed by the spin-waves,
does not form a localized
state but simply increases its effective mass.
In other words, the hole wavefunction retains a finite overlap
($Z_{\bf k}$) with the
bare (Bloch) hole state. Hence, the single hole spectral density
exhibits a quasiparticle-like (QP) band that reflects the coherent
propagation of the hole. Besides this coherent band, some
interesting features appear in the spectral density at higher energy
that can be interpreted as an incoherent background\refto{dyna,BR,scaling2}
and/or as being reminiscent of some string
levels.\refto{spectralana,spectralnum}
Basically a very similar
picture survives when quantum fluctuations are turned off
(Ising limit, $J_\perp=0$)
although with larger effective masses.\refto{barnes,trugman}

The generic t--$J_z$--$J_\perp$ Hamiltonian is,
$$
{\cal H} = {\cal T}+J_z \sum_{{\bf i},{\vec \epsilon}} S^z_{\bf i} S^z_{\bf
i+\vec \epsilon}
+ \frac1{2}J_\perp \sum_{{\bf i},{\vec \epsilon}} (S^+_{\bf i} S^-_{\bf
i+\vec \epsilon} + S^-_{\bf i} S^+_{\bf i+\vec \epsilon} ),
\tag{tj}
$$
\noindent
where $c_{{\bf i},\sigma}$ are $hole$ operators;
the hole kinetic term reads,
${\cal T}=- t \sum_{{\bf i},{\vec \epsilon},{\sigma}}
(c^{\dagger}_{{\bf i},{\sigma}} c_{{{\bf i}+{\vec
\epsilon}},{\sigma}} + h.c. ),$  and
the sum over ${\bf i},{\vec\epsilon}$ is restricted
to nearest neighbor bonds along $\vec x$ and $\vec y$ on a 2D square lattice.
We have explicitly separated the diagonal antiferromagnetic coupling
($J_z$) from the exchange one ($J_\perp$).
We shall consider two important special
cases: (i) $J_\perp =0$ i.e. the Ising limit (no quantum fluctuations), and
(ii) $J_\perp=J_z$ i.e. the isotropic case (Heisenberg).
The low-energy physics of the high-$T_c$ materials
(where the spin-orbit coupling is very weak) is well described by
$J_\perp=J_z\sim 0.3$ (if t is set to 1).\refto{tvalue}

Finite size corrections
may play an important role on small cluster calculations.
It is thus interesting to extend previous work to larger clusters.
As will be seen, from the variation of physical properties with
system size it is possible to obtain fairly well--defined results in the bulk
limit.
In this paper, we show ED results for small 2D clusters of increasing size
$N$, but with a fixed hole number $N_h=1$. Our aim is to extrapolate
various quantities to the thermodynamic limit $N\rightarrow\infty$ (that
corresponds to a vanishing hole density). Even if no simple scaling law exists,
we shall be able to give good estimates
for the extrapolated values (or in some cases rigorous upper or
lower bounds). Periodic boundary conditions (BC)
are used and we restrict ourselves to clusters of square geometry
in order to obtain a smooth behavior with increasing system size.
The cluster shape is determined by two orthogonal translation vectors
${\bf T}_1=(n,m)$ and ${\bf T}_2=(-m,n)$, where n and m can be any integers,
and the number of non-equivalent sites is then $N=n^2+m^2$. It is also required
that $N$ be even to avoid frustration of the antiferromagnetic spin background.
In other words, the antiferromagnetic wave vector ${\bf Q}_0=(\pi,\pi)$ should
always belong to the (discrete) reciprocal space.
According to these rules, cluster sizes that can be considered contain
$N=$ 8, 10, 16, 18, 20, 26, 32, ... sites. $N=8$ is too small to be
of interest, while $N=32$
is too large to be exactly handled by present-day computers.\refto{pubelbio}
The Hamiltonian is diagonalized by a Lanczos procedure
in the $S_z=1/2$ sector.
In the isotropic case this can always be assumed because of spin rotational
invariance, while for the Ising limit it is believed that the
ground state belongs to this
class. Translation symmetries are also used to further block-diagonalize
the Hamiltonian. Of course, the grid of $\bf k$-points in the
Brillouin zone (BZ) changes with the size $N$. In particular, the ground state
momentum is
changing discontinuously with $N$.
Since we are particularly interested in ground state momenta
$\bf k$ that are not necessarily located at symmetric
points of the reciprocal lattice, in general we cannot use
all the $C_{4v}$ symmetries of the square lattice.

For the larger cluster of $N=26$ sites
the relevant Hilbert space is spanned by a basis of $\sim 5.2\times 10^6$
states. In general the Lanczos vectors are complex vectors
in this basis
so that ``real'' tables of size $\sim 10^7$
are necessary to represent
each of them. The Hamiltonian matrix, which occupies around 2 Gbytes of
disk space
is only calculated once. The basic matrix-vector multiplication of the Lanczos
procedure takes around 2 min. of CPU on a Cray-2 supercomputer
so that the complete
computation of the spectral function requires of the order of  20 hours
of CPU for each set of parameters.

\vskip 1.4cm

\centerline{\bf II. SINGLE HOLE SPECTRAL FUNCTIONS}

\centerline{\bf A. General formulas}

The single hole spectral function is defined by
$$
A_{{\bf k},\sigma}(\omega )=\frac{1}{\pi }\lim_{\epsilon\rightarrow 0}\,Im\,
\{\big<\Psi_0^{N}\mid c_{\bf k,\sigma}^\dagger
\,\frac{1}{\omega +E_0^{N}-{\cal H}+i\epsilon}\,
c_{\bf k,\sigma}\mid\Psi_0^{N}\big>\} .
\tag{spectral}
$$
$\mid\Psi_0^{N}\big>$ is the ground state at half-filling (N spins on N sites)
which can be either the classical N\'eel ground state (Ising limit) or the
quantum Heisenberg ground state (isotropic limit), while $E_0^N$ is its
corresponding
energy. It is easy to understand the physical meaning of the
hole spectral density by using the complete
set of one hole eigenstates (i.e. N-1 spins)
$\{\Psi_m^{N-1}({\bf k})\}$ of corresponding energies $E_m^{N-1}(\bf k)$,
and by rewriting the spectral density as,
$$
A_{{\bf k},\sigma}(\omega )=
\sum_m\mid\big<\Psi_m^{N-1}({\bf k})\mid c_{\bf k,\sigma}
\mid\Psi_0^{N}\big>\mid^2\delta (\omega +E_0^{N}-E_m^{N-1}(\bf k)).
\tag{lehman}
$$
\noindent
The results that follow will contain a small $\epsilon =0.02$
to give a finite width to the $\delta$-functions.
For points of symmetry in the BZ such as
${\bf k}=(0,0)$, ${\bf Q}_0$, etc ..., the selection rules
impose the constraint that the manifold
$\{\Psi_m^{N-1}({\bf k})\}$ is restricted to the most symmetric
irreducible representation of the point group (such as s-wave).
For $N\rightarrow\infty$, long range AF
order implies a doubling of the unit cell and, in particular,
$E_0^{N-1}({\bf k+Q_0})=E_0^{N-1}({\bf k})$ (${\bf Q_0}=(\pi,\pi)$).

In the present work,
the spectral densities are calculated directly on finite
clusters (16 to 26 sites) by a continued-fraction expansion (truncated
after $\sim 300$ iterations) based on the
Lanczos algorithm. The calculation is made in three steps: (i) the ground state
at
half-filling is computed in a first Lanczos run (in the Ising limit
this is simply the N\'eel configuration); (ii) the hole fermion
operator is applied to the half-filled ground state; and (iii) the latter state
is used
as the initial state of a second Lanczos run. The required
memory and disk space, as well as CPU time, increases exponentially fast with
the
system size. Hence, handling the 26 site cluster (with 5.2 millions
configurations compared to only 6435 for $N=16$) is, technically,
an important step forward which requires at least $\sim 6$ Gbytes of disk
space.

In the thermodynamic limit, while most of
the $\delta$-functions merge to form a broad continuum,
a sharp $\delta$-peak persisting at the bottom of the spectrum
will be the signature of a quasiparticle excitation. Its relative amplitude
(that shoud remain finite when $N\rightarrow\infty$)
is simply given by the matrix element of the hole operator between the
lowest one-hole eigenstate (m=0)
and the initial half-filled ground state,\refto{elbiobob}
$$
Z_{\bf k}=\sum_{\sigma}\mid\big<\Psi_0^{N-1}({\bf k})\mid c_{\bf k\sigma}
\mid\Psi_0^{N}\big>\mid^2 .
\tag{zfactor}
$$
Note that the integrated spectral weight (summed over $\sigma$ and frequency)
is equal to twice the occupation number $n_{\bf k\sigma}
=\big< c^\dagger_{{\bf k},\sigma}c_{{\bf k},\sigma}\big>_N$
for each spin, i.e. it becomes 1 at half filling.
Hence $Z_{\bf k}$ varies between 0 and 1, and represents the relative weight
located in the low energy peak with respect to the total spectral weight.

\bigskip
\centerline{\bf B. Size Dependence}

Let us first analyze the spectral function of one hole in the
anisotropic (Ising) limit.
The spectral function at zero momentum is shown in Fig. 1 for the case of
the $t-J_z$ model, at the particular coupling $J_z = 0.3$. From previous
work\refto{dyna}
we know that in this limit, where the
spin fluctuations are suppressed,
the ground state of a single hole has precisely momentum
${\bf k}=(0,0)$. In Fig. 1 we observe
that the sharp peak at the bottom of the spectrum does not
change  appreciably
its position as a function of the cluster size.
Furthermore, Table 1 (containing
data for $J_z=0.3$) shows also a rapid saturation with increasing size
of the spectral weight located in the lowest energy peak.
This indicates that even a
small $4 \times 4$ cluster produces a good qualitative
approximation to
the bulk behavior of the quasiparticle in the
$t-J_z$ model. At higher energies, the situation is somewhat
more complicated. On the $4 \times 4$ cluster there are several
``satellite''
peaks in the broad energy
range $ -1 \leq \omega \leq 1$, plus additional spectral
weight at higher energies. Increasing the lattice size we observed
that the broad structure actually
collapses into a  single sharp peak located
approximately
at $\omega \sim -1.3$ for the N=26 sites cluster.
We believe that this peak is due to the
well-known ``string'' states that have been described previously
in the literature.\refto{dyna}
For $J_z>0.5$ the results for N=20 and 26 are very
similar, and thus it is expected that increasing further the lattice
size no significant changes will be noticed in the spectral function.

For smaller $J_z$, these string states are difficult to
observe owing to the small value of the coupling. Indeed, the characteristic
cluster size $N_c(J_z)$ above which the string state becomes visible increases
significantly with decreasing $J_z$; e.g. for $J_z=0.3$, the corresponding
peak can be identified unambiguously only on the largest cluster of 26 sites
(Fig. 1d).
We remark that the appearance of the string state coincides with the
saturation of the spectral weight of the low energy peak. Table 2 shows
that this saturation occurs only for large $J_z$,
even for the case of the largest clusters $N=26$,
as it will be discussed later in more detail.
{}From Fig.1 we also note that a gap clearly develops
separating the sharp quasiparticle peak
from the rest of the spectrum. This gap is
related to the finite energy required to create a spin excitation
(spin flip).

In Fig. 2 results are presented for the spin isotropic $t-J$ model.
As previously observed in Fig. 1 for the anisotropic case,
and as will be discussed in more
detail later, the sharp ``quasiparticle'' peak at the bottom of
the spectrum is not strongly affected when the size of the cluster
is changed from N=16 to 26 (see Table 3 for details).
On the other hand, the structure at
higher energies varies considerably as we increase the size of
the lattice. Consider the structure found centered near $\omega \sim 2.2$
on the $4 \times 4$ cluster. This structure was previously observed
by Dagotto et al.\refto{dyna}
and can be simply understood from the  results in the limit
$J\rightarrow 0$. There, the
spectral function presents a strong valley at $\omega = 0$, which is
symmetric under the reflexion $\omega \rightarrow -\omega$. In
Fig. 2 we observe that increasing the size of the cluster that
feature survives and in addition it broadens. This is
quite reasonable, since states at
high energies can decay into lower energy states, and thus they will
acquire finite widths. It is interesting to observe that the N=26
cluster is large enough to produce this broadening effect.
More subtle is the situation at lower
energies. The $4 \times 4$ cluster shows a sharp peak at
$\omega \sim -0.85$ which in previous studies\refto{dyna} was associated with
the first excited state of the hole moving in a linear potential
(``string'' states), with the ground state in this potential
corresponding to the quasiparticle at the bottom of the spectrum. A study
of the
J-dependence of these peaks supported that conjecture.\refto{dyna}
In contrast, in the present study we observe that
increasing the size of the lattice, that peak shifts its position
to lower energies and actually for N=26 it is now located at
$\omega \sim
-1.65$ i.e. at a considerable distance from its position on the
$4 \times 4$ cluster.
It is reasonable to expect that the first
excited string state would acquire a finite width, but its large
change in position is difficult to explain. It is unfortunate that the
N=20 and N=26 clusters do not give very similar results contrary to
what occurs for the $t-J_z$ model, and thus it is also
not clear if N=26
is a good representative of the bulk limit. There are
two possible explanations for these puzzling results: (i) As
explained before, the momentum selected to plot
the spectral functions changes
as N increases from 16 to 26. It is expected that
one hole in an antiferromagnet has a momentum ${\bf k}=(\pi/2,\pi/2)$
and only the $4\times 4$ cluster calculation can be carried out at
that precise momentum. For the N=18 cluster,
the momentum is
${\bf k^\ast} = (\pi, \pi/3)$ i.e. it is
considerably shifted from the expected bulk $\bf k$ of a hole. However,
note
that this momentum is still close to the Fermi surface of the half-filled
free electron system where ${\bf k} = (\pi/2,\pi/2)$ and
$(0,\pi),(\pi,0)$
are degenerate. Nevertheless, it is not clear whether
the results for N=18 are a good approximation
to a hole on the Fermi surface of the model in the bulk limit.
The cluster of N=20 has
${\bf k^\ast} = (4\pi/5,2\pi/5)$, while that for N=26, ${\bf k^\ast} =
(9\pi/13,7\pi/13)$, and
thus similar comments apply in these cases. However, note that only
the $4\times 4$ cluster (which can be mapped on a four-dimensional cube)
has known extra hidden symmetries.\refto{dyna}
In this respect, larger clusters maybe
more reliable.
(ii) The other alternative is that a finite-size gap separating the peak at
the bottom of the spectrum from the higher energy structure occurs leading
to artificial structure at the edge. Physically, such a gap may be
explained by the absence of long wavelength spin excitations on a finite
cluster.
This gap is expected to decrease with increasing size
(when spin excitations with longer and longer wavelengths can be created)
and eventually
disappear in the thermodynamic limit. Such an effect could explain the transfer
of spectral weight to lower energy which is seen in Fig. 2 since it leads to
a reduction of the gap. Of course, this analysis neglects interactions
between spin-waves.

In Fig. 3 we present the spectral function for the t--J model
corresponding
to zero momentum. As was previously noticed,\refto{dyna}
for this momentum the
hole  presents an accumulation of spectral weight at large energies
near $\omega \sim 1$. The actual lowest energy state of this
subspace is at a much lower energy, but the spectral
function tells us that its overlap with a Bloch state of zero
momentum is negligible (see Table 3).
Note also that increasing the size of the
lattice, the sharp peak observed for the $4 \times 4$ cluster broadens,
although its weight remains approximately constant. As discussed
before, this result is
quite reasonable, since we do not expect to find ``stable'' sharp
states at high energies. These states will acquire a finite width
since they can decay into other states of lower energy. Then,
although the $4 \times 4$ cluster captured
the qualitative physics correctly (namely, most of the
weight concentrated at high
energies), the actual shape of the distribution was not properly
accounted for and a study on a N=26 sites cluster was necessary to
obtain this behavior. A similar situation may have occurred in previous studies
of
the Raman scattering problem using these clusters,\refto{raman} where the
moments of the distribution were in excellent agreement with
experiments and series expansions, but the shape of the spectrum
contained only one dominant sharp peak similar to that of Fig.3a.
A similar phenomenon occurs for ${\bf k}=(\pi,\pi)$ shown in Fig. 4. Again the
$4 \times 4$ cluster captured the essential physics of this
particular case, namely a sharp peak at the bottom of the spectrum
but a considerable amount of spectral weight at higher energies. The
effect of increasing the lattice size is (i) to smear
the high energy features, and (ii) to reduce the magnitude of the
gap above the lower energy peak in agreement with our discussion above.

Even though the smallest cluster clearly revealed the presence of a
sharp peak at the
bottom of the band, only the behavior of this peak with increasing
system size can help us to distinguish between two very
different possibilities:
(i) a sharp peak with a small but finite width,
and (ii) a true $\delta$-function corresponding to
a real quasiparticle peak (QP). In the first case, (i) the unique peak
of the $N=16$ cluster
would split in several sub-peaks increasing the size $N$. However, we do not
observe indications of such a behavior. Actually,
Tables 1, 2 and 3
and previous analysis\refto{scaling2} suggest that the
spectral weight $Z_{\bf k}$ remains non-zero with increasing $N$. This is
particularly
clear in the Ising limit for large $J_z$ (Table 2), or in the isotropic case
for momenta $(0,0)$ and $(\pi,\pi)$. Our data strongly suggest that
this is also true for momenta close to the Fermi surface.
Alternatively one can define the average of $Z_{\bf k}$ over the BZ,
a quantity which
is meaningful even in the thermodynamic limit, but which {\it should not}
be considered as an approximation for the $Z_{\bf k}$-factor at
the Fermi surface.
{}From Table 3, we observe that this quantity shows a smooth behavior
as a function of system size (for $N\ge 18$).
All these data
for the isotropic model are in very good
agreement with the self-consistent Born approximation.\refto{klr}
Indeed, both the numerical and the analytic approaches exhibit the
following similar features: (i) a strong momentum dependence of $Z_{\bf k}$,
and (ii) a behavior $Z_{\bf k} \sim a J^\nu$ at small J,
where $\nu$
is strongly momentum dependent. Our data suggest that for momenta
close to $(\pi/2,\pi/2)$, the constants in the fit are
$a\sim 0.622$ and $\nu\sim 0.598$
in good agreement with the estimates $a\sim 0.62$ and $\nu\sim 0.667$
in the SCBA, and the results of the $4 \times 4$ cluster, $\nu\sim
0.50$.\refto{dyna}
For momentum closer to the center of the zone, the two approaches predict a
rapid
decay of $Z_{\bf k}$ with decreasing J where $\nu > 2$.

At this stage it should be pointed out that Parola and Sorella
using Quantum Monte Carlo
(QMC) techniques\refto{sorella} reached a different conclusion
for the Hubbard model at $U/t =4$, namely that $Z_{\bf k} \rightarrow 0$ in
the bulk limit.
However, these calculations were performed with a
small but still non-zero magnetic field i.e. in a polarized (e.g. non-minimal
spin) state that could bias the results. Moreover, QMC requires multiple
successive extrapolations. Clearly, more work is needed
in the QMC approach.

\bigskip
\centerline{\bf C. Momentum Dependence}

Here we continue the analysis of the momentum dependence of our
results, which was
initiated in section II.B.
First note that in the thermodynamic limit of the $t-J_z$ model,
the long range antiferromagnetic order
implies that the spectral functions at momenta $\bf k$ and $\bf k+\bf Q_0$
are the same. This is already almost fulfilled for these
finite clusters provided that the coupling $J_z$ and the size are not too
small.
Consider Fig.5 which corresponds to
the $t-J_z$ model on a cluster of N=20, and $J_z = 0.5$.
For this
coupling,
the quasiparticle is very heavy, and thus there is a weak energy
dependence in the position of the first peak
in the spectrum (flat band). The
higher energy states are more interesting. For ${\bf k}=(0,0)$ (which
approximately
produces the same spectral function as
$\bf k=(\pi,\pi)$), a sharp peak is located
near $\omega \sim -0.6$ which can be explained with the concept of
string states, as discussed before.
Moving away from the momentum corresponding to the
bottom of the band, the sharp peak rapidly disappears, but
a feature, now with a finite width,
can still be observed in approximately the same
position. Note that such a
behavior cannot
be completely understood within the string scenario which simply
predicts localized
levels, and hence no  momentum dependence.

The one-hole ground state energies and the quasiparticle weights in the
20 and 26 site clusters are shown in Table 4 and 5, respectively, in the
case J=0.25.
As expected, the energy is the lowest at momenta close to $(\pi/2,\pi/2)$
and reaches a maximum at $(0,0)$ and $(\pi,\pi)$. This is consistent with
the formation at small doping of hole pockets at $(\pm\pi/2\pm\pi/2)$.
Note that the one-hole ground state energy is defined as the location of the
lowest
energy peak in the spectral function. For a special momentum (such as $(0,0)$)
it may happen that the corresponding state is not the true ground state in that
momentum subspace. The true ground state, in this case, has in fact a different
symmetry and a (slightly) lower energy.
However, only the states appearing in the spectral function can be
observed in photoemission experiments.
As seen also in Tables 4 and 5,
the weight $Z_{\bf k}$ exhibits a strong momentum dependence. The sets of data
for 20 and 26 sites seem to be consistent with each other and show
(i) a very small weight at momentum $(0,0)$ (which is, however, almost
size independent), (ii) a maximum weight around momentum $(\pi,0)$.
In the vicinity of $(\pi/2,\pi/2)$ $Z_{\bf k}$ takes intermediate values.
As noticed before these feature are properly accounted for in the
SCBA.\refto{klr}

In addition to $Z_{\bf k}$, the whole
spectral function $A_{\bf k}$ is also qualitatively very different at
the center of the zone, at $(\pi,\pi)$, and near the FS.
In the last case, $\bf k\sim \bf k^*$, the spectral function exhibits
two separated features around $\omega\sim 2.5$ and $\omega\sim -0.5$ in Fig. 2.
As discussed above, the high energy part is reminiscent of the upper edge in
the
$J\rightarrow 0$ limit and the lower energy part might be related to
a string state (although there is no conclusive evidence supporting
this interpretation). In contrast,
the spectral functions at momenta $(0,0)$ (Fig. 3) and $(\pi,\pi)$ (Fig. 4)
are qualitatively very different from those at the FS,
and the simple string picture is not enough to
explain such a difference.\refto{laughlin} The spectral function
at zero momentum in Fig. 3a (N=16) basically shows a single peak at
intermediate
energy, where most of the weight is located. When we increase the size,
this peak broadens and eventually only a broad feature ($\Delta\omega\sim 1.5$)
remains in the case N=26. We see on this example that even a limited increase
of
the cluster area (less than a factor 2) has dramatic consequences for a line
shape. Note, that this rapid broadening does {\it not} take place
for the lowest energy peaks (QP) discussed above. In the case of momentum
$(\pi,\pi)$ (Fig. 4) a broad feature with a few significant
peaks already exists in the smallest $4\times 4$ cluster. In the largest
cluster the largest peaks are more smeared, and the structure
has become slightly broader. We also note that, here again, the
finite size gap between the QP peak and the higher energy part has
shrunk considerably from 16 to 26 sites.

\bigskip
\centerline{\bf D. Density of States}

Experimentally,
the single hole density of state (per spin) defined as
$$
N_\sigma(\omega)
=\frac1{N}\sum_{\bf k}A_{{\bf k},\sigma}(\omega),
\tag{density}
$$
\noindent can be measured using photoemission (PES) and
inverse photoemission (IPES) spectroscopy, and it would be
interesting to compare theoretical predictions for
the $t-J$ model with those experiments.
Fig.6 shows the density of states for the $t-J_z$ model as a function
of lattice size, at $J_z = 0.3$. It is clear from the figure that there
is a good
agreement between the different clusters, showing that in this
anisotropic
limit finite size effects are very small, as also observed
previously in the individual spectral functions. The quasiparticle band is
clearly seen for all clusters, and some of its features will be
discussed
below. On the other hand,
in the isotropic case (shown in Fig.7), the
size dependence is somewhat stronger, but
still the qualitative physics described by the $4 \times 4$ cluster
persists on the N=26 cluster. It is clear, however, that most of the
excited states of the smallest cluster which appear as sharp peaks, have
acquired a finite width on the largest cluster making the
``incoherent'' part of the spectrum smoother.
We are certainly aware that clusters of different
sizes have different sets of discrete momenta, and that this effect
complicates the study of the bulk limit.
However,
due to the sum over
$\bf k$ this effect is averaged producing a smooth function.
It has been explicitly checked that our results fulfill
the sum rule $\sum_\sigma\int d\omega\, N_\sigma (\omega)=1$ since
$n_{\bf k} = 1/2$ in the
insulating ground state at half-filling (the absence
of a discontinuity in $n_{\bf k}$ is characteristic of an insulating phase).

As discussed before, increasing the value of $J_z$ in the
Ising limit, a gap develops between the quasiparticle band and the
rest of the spectrum. This is clearly observed in Fig.8 where
results for the N=20 cluster are plotted as a function of $J_z$.
As mentioned before, this gap is stable (and even gets slightly larger)
with increasing system size.
The explanation is simple; for very large $J_z$ the first
excitations in the spectral function will correspond to a spin
flipped in the vicinity of the hole, and we know that
a finite energy is required for this process
($\omega\sim 3J_z/2$). As shown in
Ref. \cite{dyna}, this is equivalent to moving
the hole one lattice spacing from its original position,
and thus this excitation is simply
a ``string'' of length one. The opening of a gap is thus naturally
explained by considering a large value of $J_z$.
However in this limit this excitation has a very small
relative weight $\sim t/Jz <<1$. Its weight is maximal at $J_z\sim 1$.
All these results were previously noticed
using the $4 \times 4$ cluster\refto{dyna} (although
only at sufficiently large $J_z$), and it is important to see that
they survive our finite size scaling analysis. The present study shows that
for larger and larger clusters the essential features appear also for
smaller and smaller values of $J_z$.\refto{scaling1,scaling2}

The situation for the isotropic
case, $t-J$ model, is qualitatively different and the interpretation
of Fig. 9 requires some caution.
Fig. 9 shows also at first sight that a gap develops with increasing J.
However, only further work can decide whether this gap is real.
Our previous analysis seemed to suggest that,
contrary to the Ising case, the gap collapses in the thermodynamic limit
(Fig. 2). Hence, Fig. 9 can be interpreted simply as an increase of the
{\it finite size}
gap with $J$. Indeed, as in a pure spin system or with small doping,
one would expect a finite size gap in every sector of the spin S.
More precisely, in the case of the pure Heisenberg model, the relative
excitation energy above the ground state is given by $E(S)\sim
\frac{J}{N\kappa}S(S+1)$,
where $\kappa$ is a geometrical factor that depends on the cluster
shape.\refto{tim} The case S=1 simply gives the usual finite size spin-wave
gap. In the present case, one can construct excitations above the ground state
of the
same symmetry (i.e. with $\Delta S=0$, $\Delta\bf k=0$ ..etc..)
by combining higher spin/momentum excitations which have gaps.
In the case of the spectral function, the minimum excitation would involve
at least one spin wave, so that, very crudely,
we may expect the same J-behavior of the gap as in the Heisenberg model.

\bigskip
\centerline{\bf F. Static limit t=0}

It is instructive to study the ``static'' limit $t \rightarrow 0$,
where the hole is injected at a given site and remains there. The
spins around this static hole will be affected by its presence and
this alters
the antiferromagnetic order parameter nearby. On Table 6
we present the $Z_{\bf k}$ factor and the gap between the ground
state and the first state observed in the spectral function. For
a static hole, there is no momentum dependence in the results, and
thus any ${\bf k}$ can be considered in the definition of $Z$.
We observed that $Z$, extrapolated using a $1/N$ dependence,
seems to converge in the bulk to a large number around
$\sim 0.92$. This result is in good qualitative agreement with recent
calculations by Mal`shukov and Mahan,\refto{mahan} where in
the static limit it was found $Z \sim 0.82$.
As regards the gap, its behavior is more erratic with the system
size and it is difficult to make an accurate extrapolation.
Thus, it is not clear whether the gap will survive the
thermodynamic limit, although its large values suggest
that it may remain finite.

\bigskip
\centerline{\bf G. Bandwidth}

Here, let us consider
the bandwidth of the quasiparticle. This quantity
is useful to estimate the effective mass of the carriers. It is defined
through a study of the lowest energy peak of the one-hole spectral
function for all possible
values of the momentum. The bandwidth, $W$, is defined as
the difference between the maximum and
minimum energies obtained by this procedure. The results are shown
in Fig.10. The shape of the four curves is very similar. Clearly, a
satisfactory convergence to the bulk limit has been achieved using the
N=26 cluster. For small $J$, more
specifically between $0.1$ and $0.4$, $W$ depends
linearly on $J$, while at larger couplings a relation $W \sim
J^\alpha$ with
$\alpha < 1$ fits the data better. This is very similar to what
has been observed before on smaller lattices.\refto{dyna}

\vskip 1.5cm

\centerline{\bf III. OPTICAL PROPERTIES}

In section II, the single-hole spectral function was shown to
exhibit significant weight within an energy of a few $t$ above the ground
state.
Optical transitions from the ground state to those excited
states are thus expected. We therefore devote this section to the investigation
of the optical conductivity.
Experimentally, a large ``mid-infrared'' absorption has been observed in the
high-$T_c$ copper oxides such as doped
$La_{2-x}Sr_xCuO_{4-\delta}$\refto{LaCuO4} (p-type) and
$Pr_{2-x}Ce_xCuO_{4-\delta}$\refto{ntype} (n-type).
In $YBa_2Cu_3O_{7-\delta}$
it is not yet clear whether this absorption is due to the chains
rather than the plains.\refto{123chains} Theoretically, numerical work
on the 2D t-J model \refto{conductivity,didierdrude1,didierdrude2}
reveals that this model offers a good basis to understand such a
large absorption. These investigations have been restricted so far to small
$4\times 4$ clusters, even though the role of the boundary conditions was
carefully studied\refto{didierdrude2}. It is then important to analyze the
behavior
of the optical conductivity for larger cluster sizes, in order to separate
genuine features from finite size artifacts.

\vskip 1.3cm
\centerline{\bf A. General Equations}

For convenience, the optical absorption is decomposed as
$\sigma(\omega)=2\pi D\, \delta(\omega) + \sigma^{reg}(\omega)$,
where
$$\quad \sigma^{reg}(\omega) = \frac1{2}\sum_\alpha
\sigma^{reg}_{\alpha\alpha}(\omega),
\tag{conductivity}
$$
\noindent
and $\sigma^{reg}_{\alpha\alpha}(\omega)$ is, as usual, the dynamical
current-current correlation function  in the ground state
$\vert\Psi_0\big>$ divided by the frequency.\refto{kubo}
The $\delta-$function corresponds to the Drude peak.
Note that a summation has been performed over
the two spatial directions ($\alpha$= x or y)
in order to ensure complete rotational invariance in the case
of ground state degeneracies (that frequently occurs in the
t--J model on finite clusters).
Such a dynamical correlation function is easily obtained
using
a continued--fraction expansion\refto{contfrac} generated by a Lanczos--like
algorithm.
The Drude coefficient D is simply related to the stiffness
under a twist in the
boundary conditions (BC)\refto{kohn}
of the system and can alternatively be expressed as,
$$
D=\frac{\vert\big< \Psi_0^{N-1}({\bf k^*})\mid{\cal T}
\mid\Psi_0^{N-1}({\bf k^*})\big>\vert}{4}-\frac{1}{\pi}
\int_0^\infty d\omega\, \sigma^{reg}(\omega),
\tag{drude}
$$
\noindent
where ${\cal T}$ is the hole kinetic energy (see Eq. (1.1)).
The well-known sum rule\refto{kohn} follows directly from Eq. (3.2).

Several numerical studies have investigated
the optical conductivity,\refto{conductivity} and the
Drude weight. Early exact
diagonalizations restricted to the $N=4\times 4$ cluster revealed many
interesting features in $\sigma^{reg}$, in particular some indications
in favor of the so-called Mid--Infrared Absorption Band (MAB).
However, Eq. (3.2) leads, in the case of the 16--site cluster
and in the isotropic limit ($J_\perp=J_z$) with one hole,
to negative values for D.
However, more physical results can be obtained by averaging over BC.
In fact, it is well established that the $4\times 4$ cluster (with periodic
BC) suffers from hidden symmetries that lead to extra degeneracies
and spurious effects.
Therefore, it becomes important to perform a systematic comparison between
clusters of increasing size, that we loosely shall refer to as finite
size scaling.
This is in spirit very similar to previous work
dedicated to the scaling behavior of the single hole ground state
energy\refto{scaling1}
or spectral density (see \ref{scaling2} and section II) of the t--J model.

\bigskip
\centerline{\bf B. Ising limit}

Figs. 11 and 12 show the evolution of $\sigma^{reg}(\omega)$ with increasing
system size (from 16 to 26 sites) in the $t-J_z$ model for $J_z=0.3$ and
0.5 respectively.
First, we note the presence of a gap in the absorption
(the magnitude of which is approximately independent of the system size)
and the appearance,
above this gap, with increasing system size
of a peak at $\omega\sim 1.5 J_z$.Note that the relative amplitude of this peak
rapidly decreases with decreasing $J_z$ (for $J_z<0.15$ this peak is visible
only on the N=26 cluster).
A second absorption edge is also seen at higher energy (around
$2.3 J_z$) for intermediate $J_z$ values. These features can be qualitatively
understood in the large--$J_z$ regime where a perturbation expansion
in $1/J_z$ is possible.
For $J_z\rightarrow\infty$, the current operator
moving the hole one lattice spacing in the AF background leads to a frustration
energy of exactly $3J_z/2$. Hence, it is straightforward to show that, in
this  limit, $\sigma(\omega)=\frac{4\pi}{3}(t^2/J_z)\,
\delta(\omega-\frac{3}{2}J_z)$. When $t$ is turned on,
because of delocalization of the hole additional structures appear at higher
energies, especially at $\omega\sim \frac{5}{2}J_z$.
In the other limit, for small
$J_z$, we observe strong finite size effects.
Indeed, as usual in the Ising case,
a reasonable saturation occurs at a characteristic size $N_c(J_z)$ which
exceeds
N=16 for $J_z<0.75$ and N=26 for $J_z<0.3$ (see below for details).
When $J_z\rightarrow 0$, the higher energy absorption band gets more weight
and becomes very broad with a width $\sim t$. It is
centered at frequency
$\sim 0.7t$ with a mean-value of the distribution around $1.2t$.
On the other hand, the low energy peak (mentioned above) at
frequency $\omega\sim 1.5 J_z$ becomes very small and is not visible in the
case of the smaller clusters.

The Drude weight and the total optical absorption in the $t-J_z$ model
are shown on Fig. 13a. The behavior of these quantities with the system size is
very similar to the behavior of the ground state energy\refto{scaling1} or
of the quasiparticle weight\refto{scaling2}. Indeed, saturation starts
to occur when the system size exceeds the characteristic size
$N_c(J_z)\propto J_z^{-2/3}$ of the hole wave function\refto{spectralana}.
This is clearly demonstrated in Fig. 13b where the data are plotted versus a
reduced parameter $\propto N_c(J_z)/N$. In these units, we observe an
almost universal behavior with a rapid drop of the Drude weight
for $N>N_c$. We cannot exclude the possibility that D vanishes in the
thermodynamic limit. However, our data more likely suggest that D is very
small for all $J_z$ but still finite, typically
$D<0.05$ (the maximum value of 1 corresponds to the ferromagnetic state)
i.e. an optical mass 1/D of the order of at least $20t$.
This is consistent with the observed very small value of the
bandwidth.\refto{barnes,dyna,trugman,scaling1} For large $J_z$, the
Drude weight, as the
bandwith,
seems to behave as $t^4/J_z^3$. In that case, the absorption is reduced to
$\sim\vert\big<{\cal T}\big>\vert/4\sim \frac{4}{3}t^2/J_z$
(see Eq. (3.2)).

\bigskip
\centerline{\bf C. Isotropic case}

The analysis of the isotropic case is complicated by the fact that the ground
state
momentum is finite and changes discontinuously from one cluster size to the
other. However, by averaging over spatial directions in Eq. (3.1)
we overcome the problem of the ground state degeneracy.
This procedure is formally
equivalent to averaging over the equivalent ground states.
Note that the $4\times 4$ cluster with periodic boundary conditions
gives a negative Drude weight which is unphysical.
Hence, from now on,
we shall only consider the largest clusters of N=18, 20 and 26 sites.
Fig. 14 shows the corresponding optical conductivities calculated
for $J/t=0.3$. Note that only the regular part is shown and that the Drude
$\delta$-function at zero frequency is omitted. It is important to notice that
many features seems to be maintained when the size is increased; for example
the shape of the absorption band as well as the position of the maximum
absorption ($\omega \sim 0.7$) is preserved. However, owing to the
difference in momenta for the various data, it is difficult to tell
whether the charge excitation gap observed in Fig. 14 is a finite size gap
or is real.

The data obtained for the largest clusters agree well with the data of
Ref. \cite{didierdrude2} obtained
for the small $4\times 4$ cluster by averaging over the boundary conditions.
This establishes the efficiency of the averaging procedure introduced in
Ref. \cite{didierdrude2}. It is important
that the present finite size study also indicates that the optical mass
(not very much J dependent) seems to differ from the band mass ($\sim 1/J$)
inversely proportional to the bandwidth of Fig. 10. This suggests that
vertex corrections are essential.

A comparison between the Drude weight (inverse optical mass) and
the absortion weight is given in Fig. 15. Around $J=$ $0.2$ - $0.3$,
where the results seem to be weakly size dependent we predict an optical mass
between 1.5 and 2.5. In any case, there is no doubt that, in the whole
J region studied here, the optical mass in the isotropic limit
is much smaller than in the Ising limit discussed above. For larger J
we observe a strong size dependence so that any quantitative prediction
is impossible. However, from the monotonic behaviors of the weights as a
function of system size, we speculate that in the range, let us say, $0.4<J<2$
our data gives a lower bound for the Drude weight ($\sim 0.3$) i.e. a higher
bound for the optical mass of the order of 3. This would indicate that, in that
range, the optical mass and the finite frequency absorption are not as large
as previously thought.

\head{\bf SUMMARY}
\taghead{4.}

A finite size scaling analysis of the spectral function and
of the optical conductivity
of a single hole moving in an antiferromagnetic background
has been performed on square clusters up to 26 sites.
In the Ising limit, saturation occurs when the size exceeds the
characteristic
size $N_c(J_z)$ of the spin polaron that increases rapidly
 with decreasing $J_z$ (like
$\sim J_z^{-2/3}$).
Typically, the finite-size analysis provides reliable
results for $J_z>0.3$.
In particular, the quasiparticle weight $\sim J_z^{0.86}$ is robust
with increasing size.
The spectral function depends strongly on the
momentum. At momenta
close to $(0,0)$ a feature resembling a string excitation is seen.
Due to the spin excitation gap, a small gap $\propto J_z$
separates the low energy
delta-function from the rest of the spectrum.
In the optical conductivity
an absorption peak above the gap develops with increasing $J_z$.
Both optical and band masses are shown to be very large ($>20$).

In the isotropic t--J model the situation seems very different
First, finite size gaps in the spectral functions seem to
disappear with increasing system size as one would expect
in the case of long range AF order (with long wavelength
Goldstone modes).
It is hard to reconcile the large momentum dependence of the quasiparticle
weight wih a string picture. In contrast, this dependence
is in good agreement with the self-consistent Born approximation.
An accurate estimation of the higher energy part of the spectral
functions of the t--J model is specially reliable at
momenta close to $(0,0)$ or $(\pi,\pi)$;
it consists of a broad featureless band.
On the other hand, the accuracy of the
calculation at momenta near $(\pi/2,\pi/2)$ is limited by the differing
geometries of the finite clusters.
In that case, no excited string state is observed.
We also show that the total absorption weight at finite frequency has a strong
size dependence and {\it decreases} with increasing size (for J not too small).
This suggests that the optical mass and the optical
absorption of the t--J model may have been overestimated in the past.
This would be in qualitative agreement with the rather small mid-infrared
absorption of the $YBa_2Cu_3O_{7-\delta}$ planes.\refto{123chains}

\head{\bf ACKNOWLEDGEMENTS}
The numerical calculations were done at the Centre de Calcul Vectoriel pour
la Recherche (CCVR), Palaiseau, France. We thank the CCVR for useful assistance
during completion of this work.

\vfill\eject       
\references
\refis{raman} Elbio Dagotto and Didier Poilblanc,
Phys. Rev. {\bf B 42}, 7940 (1990).

\refis{mahan} A. G. Mal'shukov and G. D. Mahan, Phys. Rev. Lett.
{\bf 68}, 2200 (1992).

\refis{LaCuO4} S. Uchida, T. Ido, H. Takagi, T. Arima, Y. Tokura,
and S. Tajima, Phys. Rev. {\bf B 43}, 7942 (1991) and references therein.

\refis{ntype} S. L. Cooper, G. A. Thomas, J. Orenstein, D. H. Rapkine,
A. J. Millis, S.-W. Cheong, and A. S. Cooper,
Phys. Rev. {\bf B 41}, 11065 (1990).

\refis{123chains}
G. A. Thomas, J. Orenstein, D.H. Rapkine, M. Capizzi, A. J.
Millis, L. F. Schneemeyer, and J. W. Waszczak, Phys. Rev. Lett. {\bf 61},
1313 (1988);
S. L. Cooper, G. A. Thomas, J. Orenstein, D.H. Rapkine, M. Capizzi,
T. Timusk, A. J.
Millis, L. F. Schneemeyer, and J. W. Waszczak,
Phys. Rev. {\bf B 40}, 11358 (1989).

\refis{elbiobob} E. Dagotto and J.R. Schrieffer, Phys. Rev. {\bf B
43}, 8705 (1991).

\refis{dyna} E. Dagotto, R. Joynt, A. Moreo, S. Bacci and E. Gagliano,
Phys. Rev. {\bf B 41}, 9049 (1990) and references therein.

\refis{review} For a recent review see E. Dagotto, Int. J. Mod. Phys.
{\bf B 5}, 907 (1991), and references therein.

\refis{tvalue} M.S. Hybertsen, M. Schluter, and N.E. Christensen,
Phys. Rev. {\bf B 39}, 9028 (1989).

\refis{klr} C. Kane, P. Lee, and N. Read, Phys. Rev. {\bf B 39}, 6880 (1989);
G. Mart\'\i nez and P. Horsch, Phys. Rev. B {\bf 44},
317 (1991); Z. Liu and E. Manousakis, FSU-SCRI preprint.

\refis{tim} H. Neuberger and T. Ziman, Phys. Rev. {\bf B 39}, 2608 (1989).

\refis{pubelbio} Recent calculations by J. Riera and E. Dagotto, FSU-preprint,
and P. Prelov\v sek and X. Zotos (unpublished) have shown that truncating
the Hilbert space may produce accurate energies.
The method is currently being tested for the t--J model.

\refis{laughlin} However, recent calculations by A. Tikofsky and R. Laughlin,
Stanford preprint, claim to explain this behavior based on a perturbative
expansion of the gauge theory associated to the flux phase.

\refis{sorella} A. Parola and S. Sorella, ETH-TH/91-5 preprint.

\refis{varma} C.M. Varma, P.B. Littlewood, S. Schmitt-Rink, E. Abrahams,
and A. Ruckenstein, Phys. Rev. Lett. {\bf 63}, 1996 (1989).

\refis{spectralnum} K. J. von Szczepanski,
P. Horsch, W. Stephan, and M. Ziegler,
Phys. Rev. B {\bf 41}, 2017 (1990);
E. Dagotto, A. Moreo, R. Joynt, S. Bacci, and E. Gagliano,
Phys. Rev. B {\bf 41}, 2585 (1990);
C. X. Chen and H.-B. Sch\"uttler, Phys. Rev. B {\bf 41}, 8702 (1990).

\refis{scaling1} Didier Poilblanc, H. J. Schulz, and Timothy Ziman,
Phys. Rev. B {\bf 46}, 6435 (1992).

\refis{scaling2} Didier Poilblanc, H. J. Schulz, and Timothy Ziman,
Phys. Rev. B (1992), in press.

\refis{spectralana} B. I. Shraiman and E. D. Siggia, Phys. Rev.
Lett. {\bf 60}, 740 (1988).

\refis{BR} W. Brinkman and T. M. Rice, Phys. Rev. B {\bf 2}, 1324 (1970).

\refis{barnes} T. Barnes, E. Dagotto, A. Moreo, and E. S. Swanson,
Phys. Rev. B {\bf 40}, 10977 (1989)

\refis{trugman} S. A. Trugman, Phys. Rev. B {\bf 37}, 1597 (1988);
{\bf 41}, 892 (1990).

\refis{didierdrude1} D. Poilblanc and E. Dagotto, Phys. Rev. B
{\bf 44} 466 (1991).

\refis{didierdrude2} D. Poilblanc, Phys. Rev. B {\bf 44}, 9562 (1991);
D. Poilblanc, {\it Proceedings of the International Conference on
Physics in Two Dimensions}, Neuch\^atel, Helvetica Physica Acta, {\bf 65},
268 (1992); D. Poilblanc, Phys. Rev. B {\bf 45} 10775 (1992).

\refis{kubo} R. Kubo, J. Phys. Soc. Japan {\bf 12}, 570 (1957); see also
G.D. Mahan, {\it Many Particle Physics}, Plenum Press, New York (1981).

\refis{contfrac} R. Haydock, V. Heine, and M. J. Kelly, J. Phys. C{\bf 8}
2591 (1975); E. R. Gagliano and C. A. Balseiro, Phys. Rev. Lett. {\bf 59}
2999 (1987).

\refis{kohn} W. Kohn, Phys. Rev. {\bf 133}, A171 (1964); D.J. Thouless,
Physics Reports, C {\bf 13}, 94 (1974).

\refis{nagaoka} Y. Nagaoka, Phys. Rev. {\bf 147}, 392 (1966).

\refis{conductivity}
A. Moreo and E. Dagotto, {\bf B 42}, 4786 (1990);
I. Sega and P. Prelov\v sek, Phys. Rev. {\bf B 42},
892 (1990);
W. Stephan and P. Horsch, {\bf B 42}, 8736 (1990);
C.X. Chen and B. Sch\"uttler, Phys. Rev. {\bf B 43}, 3771 (1991).

\endreferences

\vfill\eject       

\vglue 2.5truecm
\centerline{TABLE 1}
\vskip 1.2truecm

$$\vcenter{\tabskip=0truecm\offinterlineskip
\def\tablerule{\noalign{\hrule}}
\halign to17truecm{\strut#& \vrule#\tabskip=0.2em plus0.2em&
  \hfil#\hfil& \vrule#&

\hfil#\hfil& \vrule#& \hfil#\hfil& \vrule#&
\hfil#\hfil& \vrule#\tabskip=0pt\cr
\tablerule
&& && && && &\cr
&& $N$ && $\bf k=0$ && $\bold k=\bf Q_0$ && $\big<\,\big>_{\bold k}$ & \cr
&& && && && &\cr
\tablerule
&& && && && &\cr
&& $16$ && $0.152\, 834$ && $0.167\, 154$ && $0.200\, 55$ & \cr
&& && && && &\cr
\tablerule
&& && && && &\cr
&& $18$ && $0.180\, 666$ && $0.183\, 577$ && $0.214\, 244$ & \cr
&& && && && &\cr
\tablerule
&& && && && &\cr
&& $20$ && $0.186\, 931$ && $0.187\, 443$ && $0.216\, 376$ & \cr
&& && && && &\cr
\tablerule
&& && && && &\cr
&& $26$ && $0.193\, 562$ && $0.193\, 563$ && $0.218\, 260$ & \cr
&& && && && &\cr
\tablerule
}}
$$

\newpage
\hsize=17truecm
\vglue 2.5truecm
\centerline{TABLE 2}
\vskip 1.2truecm

$$\vcenter{\tabskip=0truecm\offinterlineskip
\def\tablerule{\noalign{\hrule}}
\halign to17truecm{\strut#& \vrule#\tabskip=0.2em plus0.2em&
  \hfil#\hfil& \vrule#&
  \hfil#\hfil& \vrule#&
  \hfil#\hfil& \vrule#& \hfil#\hfil& \vrule#&
  \hfil#\hfil& \vrule#& \hfil#\hfil& \vrule#&
  \hfil#\hfil& \vrule#\tabskip=0pt\cr
\tablerule
&& && && && && && && &\cr
&& $J_z\rightarrow$ && $0.15$ && $0.2$ && $0.3$ && $0.5$
&& $0.75$ && $1$ & \cr
&& && && && && && && &\cr
\tablerule
&& && && && && && && &\cr
&& $N=20$ && $0.051\, 746$ && $0.112\, 453$ && $0.186\, 931$
&& $0.298\, 737$ && $0.414\, 089$ && $0.508\, 902$ & \cr
&& && && && && && && &\cr
\tablerule
&& && && && && && && &\cr
&& $N=26$ && $0.094\, 914$ && $0.132\, 177$ && $0.193\, 562$
&& $0.300\, 254$ && $0.414\, 425$ && $-$ & \cr
&& && && && && && && &\cr
\tablerule
}}
$$

\newpage
\hsize=17truecm
\vglue 2.5truecm
\centerline{TABLE 3}
\vskip 1.2truecm

$$\vcenter{\tabskip=0truecm\offinterlineskip
\def\tablerule{\noalign{\hrule}}
\halign to17truecm{\strut#& \vrule#\tabskip=0.2em plus0.2em&
\hfil#\hfil& \vrule#& \hfil#\hfil& \vrule#&
\hfil#\hfil& \vrule#& \hfil#\hfil& \vrule#&
\hfil#\hfil& \vrule#\tabskip=0pt\cr
\tablerule
&& && && && && &\cr
&& $N$ && $\bold k=(0,0)$ && $\bold k=\bold Q_0$ && $\bold k^*$
&& $\big<\,\big>_{\bold k}$ & \cr
&& && && && && &\cr
\tablerule
&& && && && && &\cr
&& $16$ && $0.005\, 236$ && $0.156\, 397$ && $0.348\, 382$
&& $0.286\, 602$ & \cr
&& && && && && &\cr
\tablerule
&& && && && && &\cr
&& $18$ && $0.005\, 215$ && $0.130\, 715$ && $0.306\, 310$
&& $0.232\, 533$ & \cr
&& && && && && &\cr
\tablerule
&& && && && && &\cr
&& $20$ && $0.005\, 161$ && $0.132\, 449$ && $0.305\, 252$
&& $0.236\, 933$ & \cr
&& && && && && &\cr
\tablerule
&& && && && && &\cr
&& $26$ && $0.005\, 309$ && $0.135\, 078$ && $0.285\, 341$
&& $0.239\, 102$ & \cr
&& && && && && &\cr
\tablerule
}}
$$

\newpage
\hsize=17truecm
\vglue 2.5truecm
\centerline{TABLE 4}
\vskip 1.2truecm

$$\vcenter{\tabskip=0truecm\offinterlineskip
\def\tablerule{\noalign{\hrule}}
\halign to17truecm{\strut#& \vrule#\tabskip=0.2em plus0.2em&
\hfil#\hfil& \vrule#& \hfil#\hfil& \vrule#&
\hfil#\hfil& \vrule#\tabskip=0pt\cr
\tablerule
&& && && &\cr
&& ${\bold k}$ && Energy && QP weight& \cr
&& && && &\cr
\tablerule
&& && && &\cr
&& $(\pi,\pi)$ && $-1.674\, 524$ && 0.132\, 449& \cr
&& && && &\cr
\tablerule
&& && && &\cr
&& $(0,0)$ && $-1.525\, 393$ && 0.005\, 161& \cr
&& && && &\cr
\tablerule
&& && && &\cr
&& $(\pi,0)$ && $-2.044\, 726$ && 0.318\, 119& \cr
&& && && &\cr
\tablerule
&& && && &\cr
&& $(3\pi/5,4\pi/5)$ && $-2.001\, 307$ && 0.274\, 387& \cr
&& && && &\cr
\tablerule
&& && && &\cr
&& $(\pi/5,3\pi/5)$ && $-2.048\, 743$ && 0.256\, 381& \cr
&& && && &\cr
\tablerule
&& && && &\cr
&& $(4\pi/5,2\pi/5)$ && $-2.118\, 732$ && 0.305\, 252& \cr
&& && && &\cr
\tablerule
&& && && &\cr
&& $(2\pi/5,\pi/5)$ && $-1.861\, 504$ && 0.155\, 182& \cr
&& && && &\cr
\tablerule
}}
$$

\newpage
\hsize=17truecm
\vglue 2.5truecm
\centerline{TABLE 5}
\vskip 1.2truecm

$$\vcenter{\tabskip=0truecm\offinterlineskip
\def\tablerule{\noalign{\hrule}}
\halign to17truecm{\strut#& \vrule#\tabskip=0.2em plus0.2em&
\hfil#\hfil& \vrule#& \hfil#\hfil& \vrule#&
\hfil#\hfil& \vrule#\tabskip=0pt\cr
\tablerule
&& && && &\cr
&& ${\bold k}$ && Energy && QP weight& \cr
&& && && &\cr
\tablerule
&& && && &\cr
&& $(\pi,\pi)$ && $-1.657\, 738$ && 0.135\, 079& \cr
&& && && &\cr
\tablerule
&& && && &\cr
&& $(0,0)$ && $-1.525\, 801$ && 0.005\, 309& \cr
&& && && &\cr
\tablerule
&& && && &\cr
&& $(8\pi/13,12\pi/13)$ && $-1.900\, 625$ && 0.250\, 391& \cr
&& && && &\cr
\tablerule
&& && && &\cr
&& $(3\pi/13,11\pi/13)$ && $-2.073\, 747$ && 0.312\, 615& \cr
&& && && &\cr
\tablerule
&& && && &\cr
&& $(9\pi/13,7\pi/13)$ && $-2.115\, 790$ && 0.285\, 341& \cr
&& && && &\cr
\tablerule
&& && && &\cr
&& $(4\pi/13,6\pi/13)$ && $-2.024\, 486$ && 0.242\, 273& \cr
&& && && &\cr
\tablerule
&& && && &\cr
&& $(10\pi/13,2\pi/13)$ && $-2.068\, 107$ && 0.308\, 665& \cr
&& && && &\cr
\tablerule
&& && && &\cr
&& $(5\pi/13,\pi/13)$ && $-1.804\, 332$ && 0.119\, 784& \cr
&& && && &\cr
\tablerule
}}
$$

\newpage
\hsize=17truecm
\vglue 2.5truecm
\centerline{TABLE 6}
\vskip 1.2truecm

$$\vcenter{\tabskip=0truecm\offinterlineskip
\def\tablerule{\noalign{\hrule}}
\halign to17truecm{\strut#& \vrule#\tabskip=0.2em plus0.2em&
\hfil#\hfil& \vrule#& \hfil#\hfil& \vrule#&
\hfil#\hfil& \vrule#\tabskip=0pt\cr
\tablerule
&& && && &\cr
&& $N$ && Energy gap && QP weight& \cr
&& && && &\cr
\tablerule
&& && && &\cr
&& $16$ && $1.960\, 623$ && 0.974\, 539& \cr
&& && && &\cr
\tablerule
&& && && &\cr
&& $18$ && $2.002\, 253$ && 0.968\, 280& \cr
&& && && &\cr
\tablerule
&& && && &\cr
&& $20$ && $1.889\, 353$ && 0.963\, 236& \cr
&& && && &\cr
\tablerule
&& && && &\cr
&& $26$ && $1.660\, 038$ && 0.950\, 487& \cr
&& && && &\cr
\tablerule
}}
$$

\vfill\eject       

\bigskip
\centerline{\bf Table Captions}
\medskip

\item{1} Quasiparticle weights $Z_{\bf k}$ in the $t-J_z$ model
for different cluster sizes $N$ and $J_z=0.3$.
${\bf k}=(0,0)$ and $(\bf k=(\pi,\pi))$ are the momenta
of the hole. The last columm
corresponds to the average of $Z_{\bf k}$ over the whole BZ.

\item{2}
Quasiparticle weights $Z_{\bf k}$ at the ground state momentum
${\bf k}=(0,0)$ in the $t-J_z$ model
for different couplings $J_z$. Data for the 20 and 26 site clusters are shown

\item{3}
Quasiparticle weights $Z_{\bf k}$ in the $t-J$ model
for different cluster sizes and $J=0.3$. Momenta $(0,0)$ and $(\pi,\pi)$
as well as ground state momenta $\bf k^\star$ are shown.
The last columm
corresponds to the average of $Z_{\bf k}$ over the whole BZ.

\item{4} $\bf k$-dependence of the one hole energy and quasiparticle weight
in the $t-J$ model for $N=20$ and $J=0.3$.

\item{5} $\bf k$-dependence of the one hole energy and quasiparticle weight
in the $t-J$ model for $N=26$ and $J=0.3$.

\item{6} Quasiparticle weights in the static limit t=0 for the various
clusters.
The gap of the first excitation above the QP is also shown.

\bigskip

\vfill\eject

\bigskip
\centerline{\bf Figure Captions}
\medskip

\item{1} Single hole spectral function
for the t--$J_z$ model at the ground state momentum
${\bf k}=(0,0)$ for various cluster sizes and $J_z=0.3$.

\item{2}
Single hole spectral function for the t--J model at the ground state momentum,
${\bf k = k^\ast}$,
for various cluster sizes and $J=0.3$. The actual values of
${\bf k^\ast}$ are discussed in the text.

\item{3}
Single hole spectral function for the t--J model at
${\bf k}=(0,0)$ for various cluster sizes and $J=0.3$.

\item{4}
Single hole spectral function for the t--J model at
$\bf k=(\pi,\pi)$ for various cluster sizes and $J=0.3$.

\item{5}
Single hole spectral function of the $N=20$ t--$J_z$ cluster
for various momenta $\bf k$ in the Brillouin zone and $J_z=0.5$.
Spectra at $\bf k$ and $\bf k+Q_0$ are averaged.

\item{6}
Density of state of the t--$J_z$ model
for various cluster sizes and $J_z=0.3$.

\item{7}
Density of state of the t--$J$ model
for various cluster sizes and $J=0.3$.

\item{8}
Density of state of the 20--site t--$J_z$ cluster
for various values of $J_z$.

\item{9}
Spectral function at the ground state momentum in the 26--site cluster of the
t--J
model for increasing J.

\item{10}
Bandwidth, $W$, as a function of J for clusters with
N=16, 18, 20 and 26 sites.

\item{11}
Behavior of the regular part of the optical conductivity of a single hole
in an Ising background with increasing system size at $J_z=0.3$.

\item{12}
Behavior of the regular part of the optical conductivity of a single hole
in an Ising background with increasing system size at $J_z=0.5$.

\item{13}
Drude coefficient D (open symbols) and integrated spectral weight
$\frac{1}{\pi}\int_0^\infty \sigma^{reg}d\omega$ (closed
symbols) in the t--$J_z$ model. Data for N = 16, 18, 20 and
26 sites (indicated on the plot) are plotted vs
(a) $J_z$ and (b) $J_z^{-2/3}\times N^{-1}$.

\item{14}
Regular part of the optical conductivity of a single hole in a quantum AF
 versus frequency.

\item{15}
Drude coefficient D (open symbols) and integrated spectral weight
$\frac{1}{\pi}\int_0^\infty \sigma^{reg}d\omega$ (closed
symbols) in the t--J model vs J.
\end